\documentclass[aps,prb,twocolumn,amsmath,amssymb,floatfix,superscriptaddress,eqsecnum]{revtex4}
\usepackage{graphicx,subfigure}

\def\LSCO{La$_{2-x}$Sr$_x$CuO$_4$}

\def\BSCCO{Bi$_2$Sr$_2$CaCu$_2$O$_{8+\delta}$}
\def\C60{A$_x$C$_{60}$}

\def\HgCu3{HgCa$_2$Cu$_3$O$_{8+y}$}
\def\HgCu4{HgBa$_2$Ca$_3$Cu$_4$O$_{10+y}$}
\def\TlCu{Tl$_2$Ba$_2$CuO$_{6+\delta}$}
\def\TlCu3{Tl$_2$Ba$_2$Ca$_2$Cu$_3$O$_{10+y}$}
\def\TlCu4{Tl$_2$Ba$_2$Ca$_3$Cu$_4$O$_{12+y}$}

\def\BiCu3{Bi$_2$Sr$_2$Ca$_{2}$Cu$_3$O$_y$}

\def\8LSCO{La$_{1.88}$Sr$_{.12}$CuO$_4$}
\def\110LNSCO{La$_{1.5}$Nd$_{0.4}$Sr$_{0.1}$CuO$_{4}$}
\def\stage4LCO{La$_{2}$CuO$_{4+\delta}$}
\def\Y248{YBa$_2$Cu$_4$O$_8$}

\def\NbSe2{NbSe$_2$}
\def\TaSe2{TaSe$_2$}
\def\TiSe2{TiSe$_2$}
\def\NaCoOH2O{Na$_{0.3}$CoO$_{2y}$H$_2$O}
\def\MgB2{MgB${}_2$}

\begin{document}

\title{Theory of the nodal nematic quantum phase transition in superconductors}
\author{Eun-Ah Kim}
\affiliation{Department of Physics, Stanford University, Stanford, California 94305, USA}
\author{Michael J. Lawler}
\affiliation{Department of Physics, University of Toronto, Toronto, ON, Canada}
\author{Paul Oreto}
\affiliation{Department of Physics, Stanford University, Stanford, California 94305, USA}
\author{Subir Sachdev}
\affiliation{Department of Physics, Harvard University, Cambridge, MA 02138, USA}
\author{Eduardo Fradkin}
\affiliation{Department of Physics, University of Illinois at Urbana-Champaign, 1110 West Green Street, Urbana, Illinois 61801-3080, USA}
\author{Steven A. Kivelson}
\affiliation{Department of Physics, Stanford University, Stanford, California 94305, USA}
\date{\today}
\begin{abstract}
We study the character of 
an Ising nematic quantum phase transition (QPT)
deep inside
a d-wave superconducting state with nodal quasiparticles in a two-dimensional  
tetragonal crystal. We find that, within a $1/N$ expansion,
the transition is continuous.   To leading order in $1/N$, 
quantum fluctuations
 enhance the dispersion anisotropy of the nodal excitations, and cause strong scattering 
which critically broadens the quasiparticle (qp) peaks in the spectral function,
 except in a  narrow wedge in momentum space near
 the Fermi surface where the qp's remain sharp. 
We also consider the possible existence
 of a nematic glass phase in the presence of weak disorder. Some possible implications for cuprate physics are also discussed.
\end{abstract}
\maketitle

 \section{Introduction}
 \label{sec:intro}
 
\begin{figure}[!b]
\includegraphics[width=.4\textwidth]{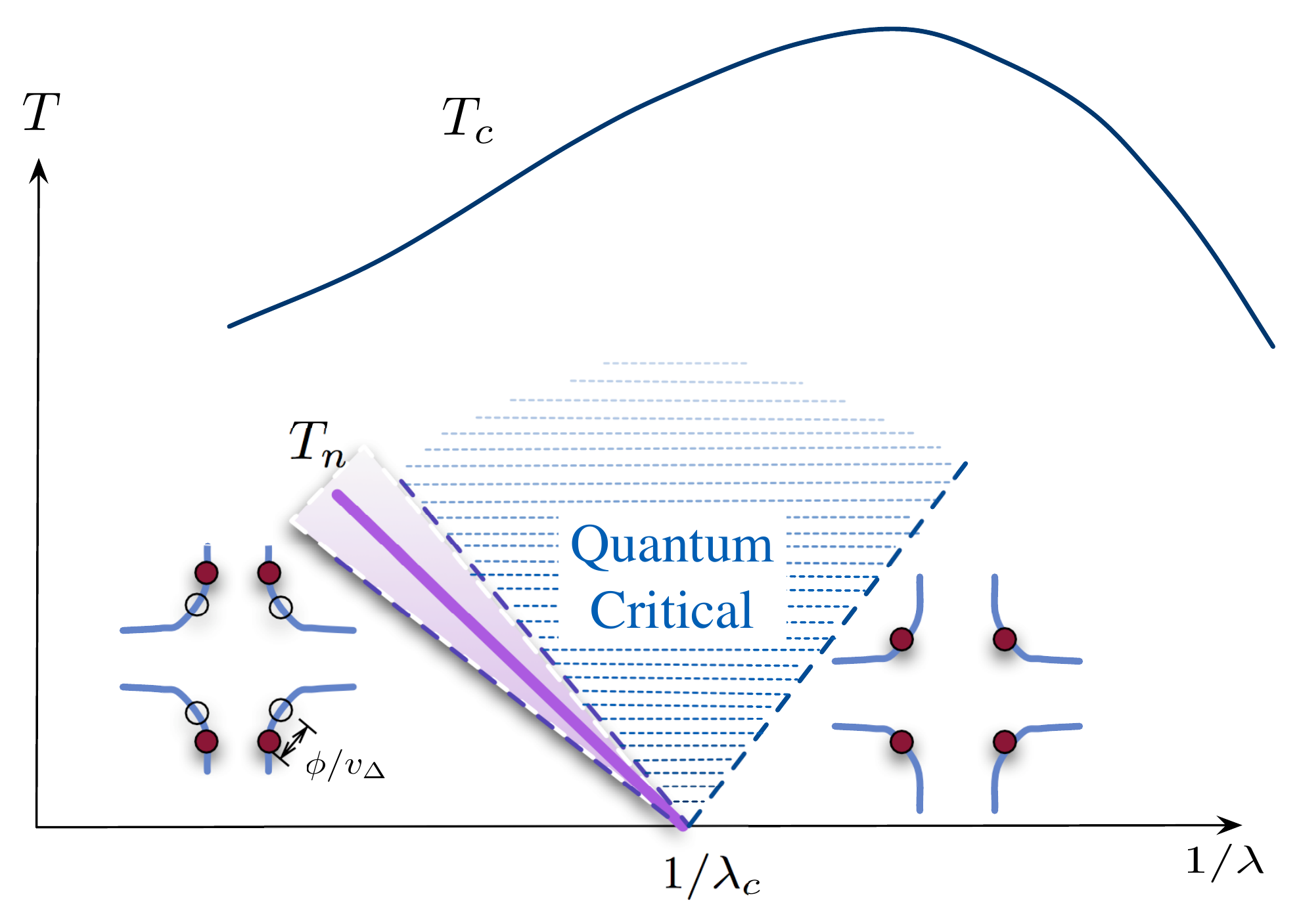}
\caption{Quantum critical point at $\lambda=\lambda_{c}$, separating the nodal nematic phase for $\lambda^{-1}<\lambda_{c}^{-1}$ from the symmetric phase, and its quantum critical fan. $T_c$ and $T_n$ are the superconducting and nematic critical temperatures and the (purple) wedge corresponds to the thermal critical regime. }
\label{fig:PD}
\end{figure}

In this paper we study the nematic to isotropic quantum phase transition (QPT), deep within the d-wave superconducting phase of a quasi two dimensional tetragonal crystal.  The ``nematic'' refers to a broken symmetry phase in which the four-fold rotational symmetry of the crystal is broken down to a two-fold symmetry.  
More specifically, it is an ``Ising nematic,'' which spontaneously breaks the discrete rotational symmetry of the tetragonal crystal to an orthorhombic subgroup, $C_{4v} \to C_{2v}$, while retaining translational symmetry.
The superconducting order opens a gap in the quasiparticle excitation spectrum, except at four gapless nodal points, but these nodal quasiparticles (qp's) couple strongly to the nematic order parameter fluctuations. In the nematic phase, these nodes are displaced from the relevant symmetry directions by an amount proportional to the nematic order parameter, as shown in Fig. \ref{fig:PD}
We derive a phenomenological theory to describe this ``nodal nematic QPT'' consisting of a nematic mode coupled to the nodal qp's of the d-wave superconductor. 

Our motivation to investigate this problem is two-fold: Firstly, there is, now, considerable experimental evidence that a nodal nematic phase occurs in at least some ``underdoped'' cuprate superconductors, so this study has potential relevance to the transition from this state to the isotropic state in these materials. The best evidence of this comes from measurements\cite{Ando2002} of strongly temperature dependent transport anisotropies in  underdoped YBa$_2$Cu$_3$O$_{6+\delta}$, and more recent (and more direct) neutron scattering experiments in underdoped YBa$_2$Cu$_3$O$_{6.45}$ (YBCO)\cite{Hinkov:2008fk}. Specifically, the spontaneous onset of one-dimensional incommensurate spin modulations in the neutron experiments is clear evidence of an isotropic-to-nematic transition with transition temperature $T_N\sim 150K$.  For $\delta=0.45$, $T_N$ is greater than the superconducting $T_c=35K$. However at higher O concentration, $\delta \gtrsim 0.7$, both the neutron scattering and transport experiments see no evidence of a transition to a nematic phase down to the lowest temperatures\cite{hinkov06,hinkov04,Ando2002,mook,stock}. It is thus reasonable to assume that $T_N(\delta) \to 0$ for a critical value, $0.45 < \delta_c < 0.7$,
with a QPT at $\delta=\delta_c$ inside the SC phase.  
\cite{kivelson1998,Vojta2000prl,Vojta2000,kivelson2003,Sachdev2003}.
The extent to which nematic phases are generic to the cuprates is a question that is beyond the scope of the present study.
However, Matsuda and coworkers\cite{matsuda-2008} have recently reported that the ``fluctuating stripe'' phase of underdoped {\LSCO} (with $x=0.04$) is actually a (diagonal) nematic phase.  Moreover, STM studies\cite{howald, Kohsaka2007} have revealed 
a glassy phase with  nematic domains in underdoped {Bi$_2$Sr$_2$CaCu$_2$O$_{8+\delta}$}.

Secondly, the topic of quantum critical points (QCP) in 2D systems with itinerant fermions 
has been and continues to be a topic of broad interest. Here we study a new QCP in a 2D itinerant fermion system. In general, the presence of gapless fermions at the Fermi surface makes the study of this topic theoretically challenging. Since gapless fermions interact with massless bosons associated with order parameter fluctuations, this interaction affects both low energy degrees of freedom, making it difficult to obtain a correct description of the 
critical physics. Simplifications occur when one considers nodal fermions which are gapless 
 only at four nodal points of the underlying Fermi surface in a d-wave superconductor. The limited phase space for the gapless Fermions restricts the possibilities for scattering mechanisms permitting a controlled analysis.

Electronic nematic phases  were first predicted to occur in doped Mott insulators\cite{kivelson1998}, and have now been experimentally observed in a number of different systems. In addition to YBCO,
they have been discovered in transport experiments in semiconductor heterostructures\cite{Cooper2002} and in the bulk transition metal oxide Sr$_3$Ru$_2$O$_7$\cite{Borzi2007}. 
While in all cases the thermal transition to the nematic phase appears to be continuous, the character of the quantum transition at zero temperature is less clear, and at least in the case of Sr$_3$Ru$_2$O$_7$\cite{Borzi2007} appears to be first order.  However, interesting universal physics can be expected in the case of 
a continuous transition.
 
Vojta, Zhang and Sachdev \cite{Vojta2000prl,Vojta2000} analyzed various types of ordering transitions in a nodal superconductor, including Ising nematic ordering. They employed an  $\epsilon$ expansion  ($\epsilon=3-d$) to derive a perturbative renormalization group (RG)  flows
near the decoupled fixed point, $\lambda=0$. In the nematic case, they found runaway flows which they tentatively interpreted as a fluctuation induced first order transition.  Of course, a runaway flow can also imply a breakdown of perturbation theory.
 
In this paper we work directly in $d=2$, and use a large $N$ theory
to access a non-trivial critical point at finite coupling $\lambda_c$. This is achieved by generalizing the problem to one in which there are $N$ ``flavors'' of nodal qp's (with the physical $N=2$ case corresponding to two spin polarizations)\cite{huh}.
In the large $N$ limit, we present a theory of the nodal nematic QPT\cite{khveshchenko},  whose critical behavior should be smoothly 
and systematically corrected by the $1/N$ expansion\cite{wilson,gross-neveu,zinn-justin,moshe}.  
We find: 
a)  the quantum phase  
transition is continuous, with non-trivial critical exponents that we compute for $N\to \infty$,
b) the transition occurs at a {\it finite} critical coupling between the nodal qp's and the nematic order parameter, so is inaccessible by perturbation theory, 
c) the anisotropy of the nodal qp dispersion strongly influences the interplay between the critical fluctuations and the qp's. We also speculate on the implications of this theory to the phenomenology of nodal qp's in cuprates. With minor changes, this theory also applies to a possible electronic hexatic QPT in graphene and similar systems.\cite{eakim-preprint}

The paper is organized as follows. In Section \ref{sec:model} we introduce the model  and discuss its symmetries. The effective theory of the nematic order parameter is derived, in the large $N$ limit, in Section \ref{sec:op-theory}. In Section \ref{sec:nodal-qcp} we discuss the behavior of the nodal quasiparticles at and near the nematic quantum critical point. In Section \ref{sec:glass} we give a qualitative presentation of of the physics of a nematic glass and how it affects the spectrum of nodal quasiparticle, leading to a Fermi arc phenomenology in the quasiparticle spectral function.  In Section \ref{sec:discussion} we discuss a number of important open problems.

\section{The model}   
\label{sec:model}

The effective Lagrangian which describes the coupling of the nodal fermionic qp's to the nematic order parameter is 
$\mathcal {L}=\mathcal{L}_{\Psi}+\mathcal{L}_{\mathrm {int}}+ \mathcal{L}_{\phi}$, where $\mathcal{L}_\Psi $
 is the linearized nodal qp Lagrangian in a pure $d_{x^{2}- y^{2}}$ SC
\begin{equation}
\mathcal{L}_\Psi =\sum_{n,\alpha}\bar{\Psi}_{n,\alpha}\left(\partial_{\mathcal {\tau}}-i\tau_3\vec{v}^{\;n}_F\cdot \vec \nabla-i\tau_1\vec{v}^{\;n}_\Delta\cdot \vec \nabla\right)\Psi_{n,\alpha}.
\label{eq:L_Psi}
\end{equation}
Here
\begin{equation}
\Psi_{n,\alpha}(\vec{p})\equiv
\begin{pmatrix}
{\phantom{\epsilon_{\alpha\beta}}}c_{\vec{K}_{n}+\vec{p},\alpha}^{\phantom\dagger} \\
\epsilon_{\alpha\beta}\ c_{-(\vec{K}_{n}+\vec{p}),\beta}^{\dagger}
\end{pmatrix}
\label{eq:Nspinor}
\end{equation}
 are two-component Nambu spinors representing the nodal qp's for node index
 $n=1,2$ and  $\alpha,\beta=1, \ldots, N$ are $Sp(N/2)$ ``flavor'' indices which,
for $N=2$, correspond to the qp  spin polarizations, since
 $Sp(1)\simeq SU(2)$.
 Eq.\eqref{eq:Nspinor} represents nodal qp's
  at momentum positions 
 $\vec{p}$ relative to each pair of nodes at
$\vec{K}_1=(K,K)$  and  
$\vec{K}_2=(-K, K)$ for $n=1,2$ respectively and
\begin{equation}
{\vec v}_F^{\;n}=\vec{\nabla}_{k}\epsilon_{k}|_{\vec{k}=\vec{K}_{n}},\quad  {\vec v}_\Delta^{\;n} = \vec{\nabla}_{k}\Delta_{k}|_{\vec{k}=\vec{K}_{n}}
\end{equation}
 are velocities normal and tangential to the Fermi surface (FS).
 Note that we investigate our model for general values of 
 $v_F$ and $v_\Delta$.
 We use $\tau_1$ and $\tau_3$ to denote  $2\times 2$ Pauli matrices acting on the Nambu spinors.
The Lagrangian 
for the nematic order parameter $\phi$ is  
\begin{equation}
\mathcal{L}_{\phi} = \frac{m^2}{2}\phi^2+
  \frac{1}{2}[(\partial_\tau\phi)^2+c^{2}(\nabla\phi)^2] + \frac{u}{4N} \phi^4+\cdots
\label{eq:Lphi}
\end{equation}
where the ellipsis 
represent higher order terms in powers of  the nematic order parameter field
 $\phi$. We assume that $u>0$ and that $m^2>0$, {\it i.e.\/}  in the absence of coupling to the nodal qp's, 
 the system is in its isotropic phase,
$\langle \phi\rangle=0$, and
the nematic mode is gapped.
The interaction term that couples the nematic order parameter to a fermion bilinear is 
\begin{equation}
\mathcal{L}_{\rm int}  =\frac {\lambda}{\sqrt{2N}} \sum_{n,\alpha}
\phi\bar{\Psi}_{n,\alpha}\tau_1\Psi_{n,\alpha},
\label{eq:Lint}
\end{equation}
which has a well defined large $N$ limit\cite{wilson,gross-neveu,zinn-justin,moshe}. 
This model has the discrete symmetry $(\phi\rightarrow -\phi, \vec{p}\rightarrow-\vec{p},\; \Psi\rightarrow\tau_{2}\Psi)$. The combination of this discrete symmetry with the implicit symmetry of the Nambu notation and the two fold rotation symmetry  associated with $(\Psi_{1}\leftrightarrow \Psi_{2},\; k_{x}\leftrightarrow k_{y})$ amounts to $C_{4v}$ symmetry.
Should $\langle \phi\rangle\neq0$, the position of the nodes shift in the $\vec{v}_\Delta^n$ direction and  the $C_{4v}$ symmetry is spontaneously broken down to  $C_{2v}$
associated with $(\Psi_{1}\leftrightarrow \Psi_{2},\; k_{x}\leftrightarrow k_{y})$ (see Fig.~\ref{fig:PD}).

The interaction in Eq.\eqref{eq:Lint} has the form of a Dirac fermion current coupled to one component of a vector gauge field. Couplings of this type have been studied in the context of gauge theoretic approaches 
 to nodal fermions and phase fluctuations\cite{balents98,Vafek2002,hermele2005}. However, in the case at hand, the microscopic theory {\em does not have a local gauge invariance} associated with the nematic field $\phi$,
but instead only has a global {\it discrete} symmetry which can be broken spontaneously.

Since the coupling makes the tangential and normal directions to the FS inequivalent, one {\it must } consider an anisotropic dispersion, $v_F/v_\Delta\neq 1$. Indeed, in previous studies of quantum criticality with Dirac-like fermions, an effective rotational symmetry about the nodal point, $v_F/v_\Delta\to 1$, was shown to be an emergent property of the fixed point in the sense that anisotropy is perturbatively irrelevant; this is not true in the present problem, and indeed we find a number of qualitatively new features of the critical theory which derive from this anisotropy!

The fact that $v_F$ and $v_\Delta$ have different physical origins, implies that there is no reason a priori to expect the existence of  
even an approximate symmetry $v_F=v_\Delta$.
Indeed, experimentally it is well known that the degree of anisotropy is extreme: $v_F/v_\Delta$ $\sim 19$ for BSCCO from thermal conductivity\cite{chiao} and ARPES\cite{norman}, and  $v_F/v_\Delta$ $\sim 14$ for YBCO from thermal conductivity\cite{chiao}. After we formulate our theory for general values of $v_F$ and $v_\Delta$, we investigate the specific case which corresponds to linearizing the phenomenological band structure of M.\@ Norman {\it et al.\ }\cite{norman}. 
  
\section{The order parameter theory} 
\label{sec:op-theory}

We will now proceed to study the behavior in the large $N$ limit of the theory of nodal fermions coupled to a nematic order parameter theory presented in the last section. As it is standard in the large $N$ limit we proceed to formally integrate out the fermions and 
obtain the resulting  effective action for the nematic order parameter field.
\cite{wilson,zinn-justin,moshe}
 In the large-$N$ limit, the resulting bosonic path integral 
 \begin{equation}
 \mathcal{Z}=\int\mathcal{D}\phi\,e^{-S_{\mathrm{eff}}[\phi]}
 \end{equation}
  will be dominated by the saddle point solution
 to the effective action $S_{\rm eff}[\phi]$,
  and is amenable to a controlled $1/N$ expansion to evaluate the effects of fluctuations. In imaginary time, the full effective action $S_{\rm eff}[\phi]$ has the standard form
  \begin{eqnarray}
 && S_{\rm eff} [\phi]=\nonumber \\
  &&\int d^3x \left[
 \frac{m^2}{2}\phi^2+
  \frac{1}{2}[(\partial_\tau\phi)^2+c^{2}(\nabla\phi)^2] + \frac{u}{4N} \phi^4\right]
  \nonumber \\
  &&-N\sum_n \ln \textrm{Det} \left(\partial_\tau -i \tau_3 {\vec v}^n_F\cdot {\vec \nabla}-i \tau_1 {\vec v}^n_\Delta\cdot {\vec \nabla}+\frac{\lambda}{\sqrt{2N}} \; \tau_1\; \phi \right)
  \nonumber \\
  &&
  \label{eq:effaction}
  \end{eqnarray}
The fermion determinant in Eq.\eqref{eq:effaction} represents the sum of fermion bubble Feynman diagrams. Since the action of the nodal fermions has the  structure of a anisotropic Dirac theory with a coupling to the nematic order parameter $\phi$, each bubble diagram involves a trace over a product of  Dirac fermion propagators with the appropriate insertions of the matrices $\tau_1$ dictated by the form of the coupling. This is a standard procedure which obviates the need to present the well known intermediate steps. We will thus only present the main results of these calculations.

Upon a rescaling the order parameter  $\phi \to \sqrt{N} {\phi}$, it is apparent that the effective action for the rescaled field ${\phi}$ has the form $ N S_{\rm eff}[ \phi]$. Hence, in the large $N$ limit the path integral is dominated by the solution to the saddle point equations for the effective action $S_{\rm eff}[ \phi]$, and its fluctuations. 
When expanded in powers of $\phi$, the contribution of the fermion determinant introduces a negative contribution to the quadratic term  that is proportional to $\lambda^2$ in the effective action $S_{\mathrm{eff}}[\phi]$. \cite{wilson,gross-neveu,zinn-justin} 
This allows for the spontaneous symmetry breaking driven by the coupling constant $\lambda$. In theories of this type \cite{wilson,zinn-justin,moshe} there is a critical coupling strength $\lambda_c$, beyond which $\phi$ gains a finite expectation value due to the spontaneous symmetry breaking. The critical value of the coupling constant, the location of the quantum phase transition, depends on the bare mass of the nematic mode and the UV cutoff. If we use a sharp UV cutoff $\Lambda$, in the large $N$ limit we obtain
 \begin{equation}
 \lambda_c^2 =[2\pi^2/\ln 2] (m^2 v_Fv_\Delta/\Lambda).
 \end{equation} 
For $\lambda^2< \lambda_c^2$ the symmetry is unbroken and $\langle\phi\rangle=0$.
However for $\lambda^2>\lambda_c^2$, a nonzero expectation value 
$\langle \phi \rangle$ is obtained by
minimizing $S_{\mathrm{eff}}[\phi]$. This can be achieved by 
solving the saddle point equation:
\begin{equation}
m^2\langle \phi\rangle=-\frac{\lambda}{\sqrt{2N}} \sum_{n,\alpha}\langle\bar\Psi_{n,\alpha}\tau_{1}\Psi_{n,\alpha}\rangle.
\label{eq:gapEQ}
\end{equation} 
For $\lambda$ close to $\lambda_c$, the solution of Eq.\eqref{eq:gapEQ} satisfies
\begin{equation}
\langle \phi\rangle\propto (\lambda-\lambda_c)^{1/2}. 
\label{OP}
\end{equation}
Thus, in the large $N$ limit, we find a continuous quantum phase transition, {\it i.e.\/} a quantum critical point, separating the weak coupling isotropic phase at $\lambda < \lambda_c$ from a strong coupling phase at $\lambda> \lambda_c$. The critical exponent of the order parameter of Eq. \eqref{OP}, $1/2$, obtained in the large $N$ limit will be corrected order by order in the $1/N$ expansion. \cite{wilson,zinn-justin} Within the $1/N$ expansion, the corrections to the value of this and other critical exponents of this theory follow from the resummation (exponentiation) of the infrared divergent $1/N$ perturbation theory. This procedure can also be viewed as a construction of a fixed-point theory (and its associated renormalization group) order-by-order in the $1/N$ expansion, a standard procedure in relativistically invariant theories of Dirac fermions. \cite{wilson,zinn-justin}

In the present problem, Lorentz invariance is explicitly absent due both to the velocity anisotropy, parametrized by an anisotropy ratio $\alpha\equiv (v_F-v_\Delta)/v_F$, 
and the form of the coupling between 
the nematic order parameter $\phi$ and the Dirac fermions. The lack of this  invariance  
 changes the physics in  dramatic ways. 
As noted above,  Vojta, Zhang and Sachdev studied the quantum critical behavior of a similar system by means of a perturbative renormalization group (RG) analysis at fixed $N$ near the critical (renormalizable) dimension $D=3+1$ using an   $\epsilon$ expansion.\cite{Vojta2000prl,Vojta2000} 
 Due to the combined effects of the velocity anisotropy and the anisotropic coupling,  they found that the perturbative RG has runaway flows, signaling the breakdown of perturbation theory. Following a standard analysis (see for instance Refs. [\onlinecite{zinn-justin,cardy}]) these authors interpreted this runaway  flow as an indication of a fluctuation induced first order transition.   Instead, in the large-$N$ theory that we present here, we find that, at fixed dimension $D=2+1$ but for very large $N$, the quantum phase transition is continuous and has a finite renormalized velocity anisotropy which plays a key role.\cite{order-of-limits} 
This behavior also contrasts with that of more nearly Lorenz invariant models, such as the QED$_3$ type models studied in Refs. [\onlinecite{balents98,Vafek2002,hermele2005}], where a small velocity anisotropy was found to 
be either irrelevant or a redundant. 

We will now proceed with the analysis of this theory in the large-$N$ limit. The leading quantum fluctuations at large $N$ are obtained by expanding the effective action to quadratic order about the saddle point (see for instance, Ref.\ [\onlinecite{eduardo}]). 
This standard procedure, when applied to the present problem which breaks Lorentz invariance, introduces 
two separate energy scales associated with any given momentum $\vec{p}$   
\begin{equation}
E_1(\vec{p})\equiv \sqrt{v_F^2p_x^2+v_\Delta^2p_y^2},\;
E_2(\vec{p})\equiv\sqrt{v_F^2p_y^2+v_\Delta^2p_x^2},
\label{eq:E1E2}
\end{equation}
where $E_1(p_x,p_y)=E_2(p_y,p_x)$.
In terms of $E_1(\vec{p})$ and $E_2(\vec{p})$, 
the Gaussian action $S^{(2)}[\varphi]$ for the fluctuations of the nematic mode $\varphi(x)$ defined by $\varphi(x)\equiv\frac{m}{\lambda_c} [\phi(x)-\langle \phi\rangle]$ takes the following form on the disordered side ($\lambda<\lambda_c$) of the critical point:
\begin{multline}
S^{(2)}[\varphi]=\int\frac{d^2p d\omega}{(2\pi)^3}\;\frac{1}{2}\; \Bigg[\left({\lambda^2_c}-{\lambda^2}\right)
+{\kappa}(\omega^2+c^{2}\vec{p}^{\;2})
\\
+\gamma
 \sqrt{\omega^2+E_1(\vec{p})} \left(1-\frac{v_\Delta^2p_y^2}{\omega^2+E_1(\vec{p})}\right)+[p_x\leftrightarrow p_y]\Bigg]|\varphi(p)|^2\; ,
\label{eq:full-Seff-varphi}
\end{multline}
where $\kappa\equiv\lambda_c/m^2$ and
$\gamma\equiv \lambda_c^2/32v_Fv_\Delta$. 
The quartic and higher order terms in fluctuations have coefficients which are\cite{zinn-justin,moshe} down by powers in $1/N$. Here we will only treat the physics to leading order in this expansion
\cite{caveat}.

We can analyze the effective action in Eq.~\eqref{eq:full-Seff-varphi} from the viewpoint of scaling theory.  
Clearly, at criticality ($\lambda=\lambda_c$), the bare dynamical terms in Eq.\eqref{eq:Lphi} 
are irrelevant, as
they have larger dimension than the non-local term generated by integrating out the nodal fermions. The scaling dimension  of the nematic field $\varphi$, as traditionally defined, is
$\textrm{dim}[\varphi]=(d+z-2+\eta)/2$, with  the space dimension $d$,  the dynamic critical exponent $z$, and the anomalous dimension $\eta$.  
In the present case, $\textrm{dim}[\varphi]=1$ and $z=1$ is inherited from the Dirac-like spectrum of the fermions.
This  can then be interpreted as a (large!) 
 anomalous dimension $\eta=1$~\cite{analogy}.
   As a result of this scaling, 
 all {\it local} interaction terms of higher order in $\varphi$ 
 are (sometimes dangerously) {\it irrelevant} operators. 
 (This is in contrast to 
a local $\varphi^4$ theory in which the $u\varphi^4$ term is relevant for $d<3$ at the Gaussian fixed point, so long as $z<2$, making the fixed point unstable.)
Non-local interactions are also generated by integrating out the fermions,
which,  presumably, lead to ${\cal O}(1/N)$ corrections to the 
various critical 
exponents.
Away from 
criticality, the correlation length scales as $\xi \sim |\lambda_c -\lambda|^{-\nu}$ with $\nu=1$.
To the extent that scaling holds, 
$T_c\sim\left(\lambda-\lambda_c\right)$ for $\lambda > \lambda_c$
 ($\nu z=1$). 
 The resulting phase diagram, with a v-shaped quantum critical fan  is sketched in Fig.~\ref{fig:PD}.
 
\begin{figure}[hbt]
\subfigure[]{
   \includegraphics[width=.3\textwidth]{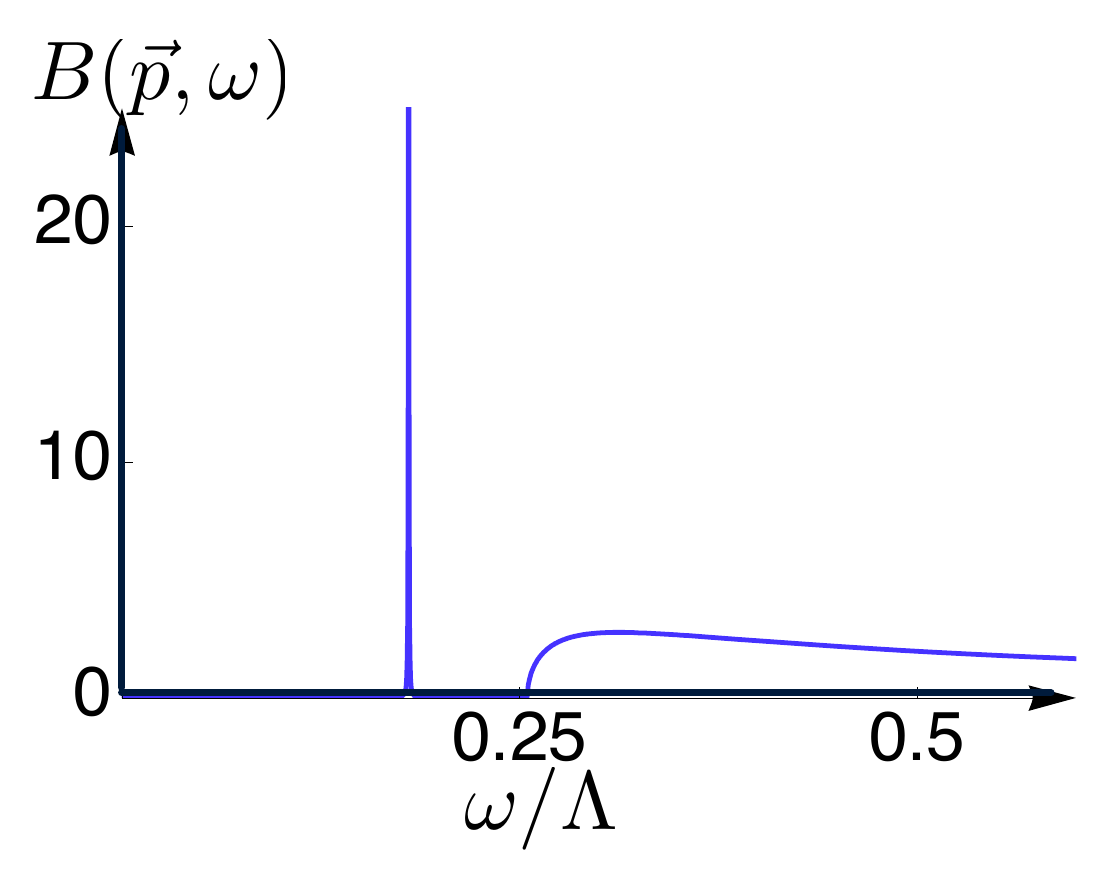} 
  }
  \subfigure[]{
    \includegraphics[width=.3\textwidth]{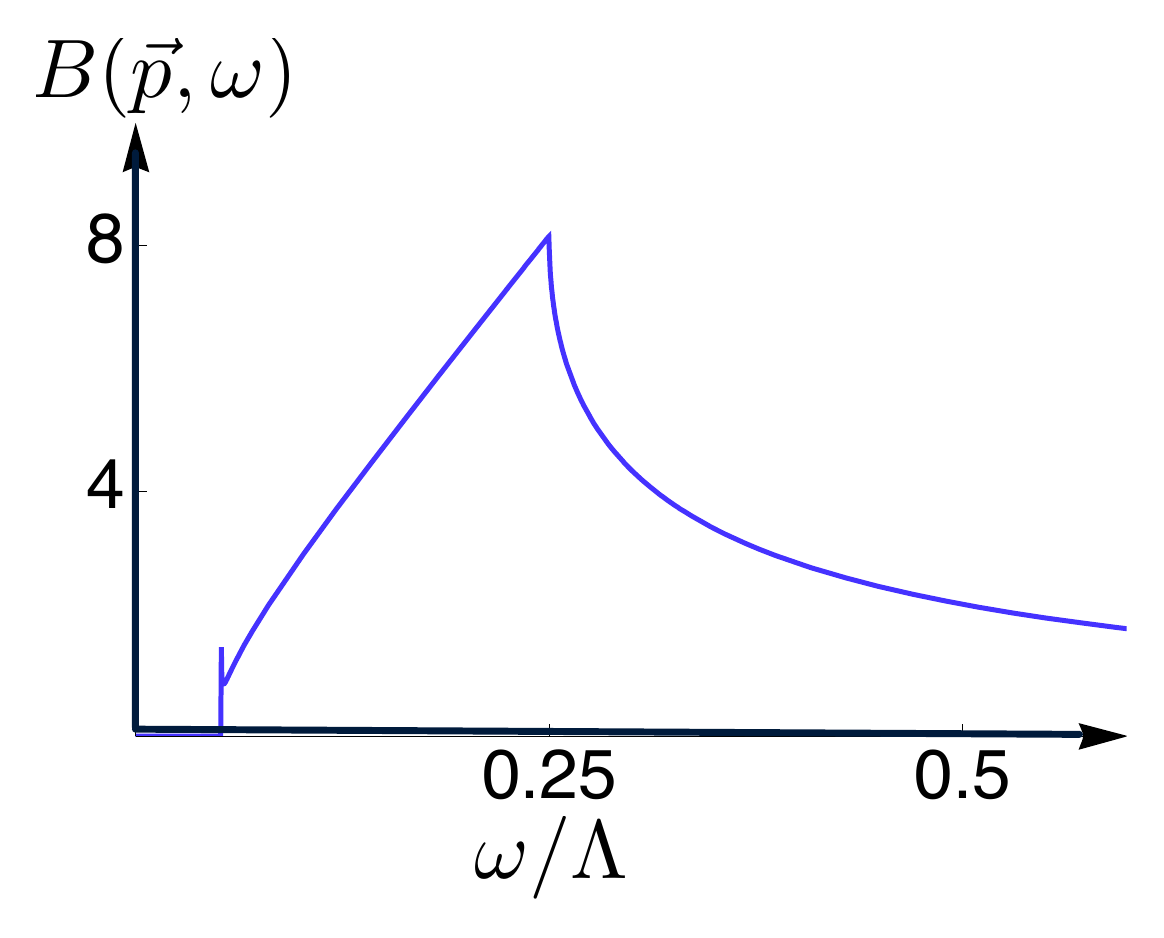}
   }   
   \caption{The nematic mode spectral function $\tilde{B}(\omega,\vec{p})$ plotted as a function of energy $\omega/\Lambda$ at momentum  
$\vec{p}=(0.25{\Lambda}/{v_F}, 0.05{\Lambda}/{v_F})$, 
where $\Lambda$ is a UV cutoff and $v_F$ is the Fermi velocity. (a) Isotropic case $v_F\!=\! v_\Delta\!=\!v$. (b) Anisotropic case $v_F/v_\Delta \!=\! 19.5$. 
}
\label{fig:B-spectral}
\end{figure}
We compute the spectral response of the nematic mode by an analytic continuation of
the effective action Eq.~\eqref{eq:full-Seff-varphi} to real time.  The irrelevant terms
in the effective action can be set to zero for energies small compared to  $\gamma/ \kappa$.  In Fig.~\ref{fig:B-spectral} we show the spectral function of the nematic mode $\varphi$ at criticality
\begin{equation}
B(\vec{p},\omega)\equiv-2{\mathrm{sgn}}(\omega)\textrm{Im}G (\vec{p},\omega),
\label{eq:B}
\end{equation}
for the $\varphi$ propagator 
\begin{equation}
{G}(\vec{p},\omega)=\big[\gamma
\sqrt{-\omega^2+E_1(\vec{p})^2}\,{\mathcal P}_1(\vec{p},\omega)+(1\leftrightarrow2)\big]^{-1},\label{eq:G}
\end{equation}
where we defined projectors
\begin{equation}
\mathcal{P}_1\!=\! \left(\!1-\frac{v_\Delta^2p_y^2}{-\omega^2\!+\!E_1(\vec{p})^2}\!\right),
\mathcal{P}_2\!=\! \left(\!1-\frac{v_\Delta^2p_x^2}{-\omega^2\!+\!E_2(\vec{p})^2}\!\right).
\end{equation}

The spectral function of Eq.\eqref{eq:B}  contains contributions from a pole and from branch cuts. 
When the dispersion is isotropic, 
the nematic mode pole is at $\omega=v|\vec{p}|/\sqrt{2}$ with residue $Z=1/4$ and there is a threshold at 
$\omega=v|\vec{p}|$ for the continuum due to the branch cut. (See Fig.\ref{fig:B-spectral}(a)\cite{eakim-preprint}) However,  the dispersion anisotropy $\alpha>0$ ($v_F> v_\Delta$) completely alters the analytic structure of the nematic mode spectral function. First of all, the pole approaches the continuum with a reduced residue\cite{eakim-preprint}. Moreover the threshold energy scale for the continuum at $\omega=v|\vec{p}|$ associated with given momentum $\vec{p}$  for an isotropic dispersion, splits into features at two energy scales $E_1(\vec{p})$ and $E_2(\vec{p})$ (See Fig.\ref{fig:B-spectral}(b)).
The spectral function $B(\vec{q},\omega)$ can in principle be 
measured with momentum resolved inelastic X-ray scattering. Observation of nematic mode spectral function with two distinct energy scales associated with a given momentum could serve as direct evidence for nodal nematic criticality. Moreover, the nontrivial analytic structure of the nematic mode spectral function results in momentum dependent scattering for quasi particles.

\section{Properties of nodal fermions at criticality}
\label{sec:nodal-qcp}
The critical nematic fluctuations have drastic effects on the nature of nodal fermions.  To illustrate this, here we will discuss the single particle spectral function at the nodal nematic QCP, defined in the vicinity of each node $n=1,2$ as 
\begin{equation}
A_n(\vec{p},\omega)\equiv-2{\rm{sgn}}(\omega)\textrm{Im} [\mathcal{G}_{n,11}(\vec{p},\omega)].
\label{eq:An}
\end{equation}
Here $\mathcal{G}_{n,11}(\vec p, \omega)$ is the $(11)$ component of the $(2\times 2)$ Nambu matrix time ordered qp propagator
\begin{equation}
\mathcal{\hat G}_n
(\vec{p},\omega)\equiv-i\langle T\Psi_{n}(-\vec{p},\omega)\bar{\Psi}_{n}(\vec{p},\omega)\rangle.
\label{eq:Gn}
\end{equation}
Note, in the rest of this section, we will focus on the vicinity of the $n=1$ node  and omit the subscript $n=1$ 
to simplify the notation.

A matrix-valued qp self-energy $\hat{\mathbf \Sigma}(\vec{p},\omega)$ best characterizes the 
effect of critical nematic fluctuations on the single fermion spectral function.
We can then express $A(\vec{p},\omega)$ in terms of $\hat{\mathbf\Sigma}$
in the standard manner.  In the presence of $\hat{\mathbf\Sigma}$, the Nambu matrix propagator for the nodal Fermions $\hat{\mathcal{G}}$ is given by $\hat{\mathcal{G}}^{-1} = \hat{\mathcal{G}}_0^{-1}-\hat{\mathbf\Sigma}$, where  $\hat{\mathcal{G}}_{0}(\vec{p},\omega)=(\omega\mathbb{I}-v_Fp_x \tau_3-v_\Delta p_y \tau_1)^{-1}$ is the Nambu matrix propagator of the free nodal Fermion theory. Decomposing the matrix valued self-energy in a basis of Pauli matrices as 
$\hat{\mathbf \Sigma}(\vec{p},\omega)\equiv \Sigma^{(0)}\mathbb{I}-\Sigma^{(1)}\tau_3+\Sigma^{(2)}\tau_1$,
the associated single particle spectral function is 
\begin{align}
&A(\vec{p},\omega)=-2{\rm{sgn}}(\omega)\times\nonumber\\
&\textrm{Im}\left[\frac{(\omega-\Sigma^{(0)})+(v_Fp_x-\Sigma^{(1)}) }{(\omega-\Sigma^{(0)})^2-(v_Fp_x-\Sigma^{(1)})^2-(v_\Delta p_y+\Sigma^{(2)})^2}\right].
\label{eq:A}
\end{align}
From Eq.\eqref{eq:A} one can understand  the effect of different components of $\hat{\mathbf\Sigma}$ on the physical properties of the fermions.

To order $1/N$, the $(2\times2)$ qp self-energy matrix at the QCP is
\begin{equation}
\hat{\mathbf \Sigma}(\vec{p},\omega)\!=\! \frac{i\lambda_c^2}{2N}\int\!\!\frac{d^2k}{(2\pi)^2}\frac{d\omega'}{2\pi}
\tau_1\hat{\mathcal{G}}_0(\vec{k},\omega')\tau_1 {G}(\vec{p}\!+\!\vec{k},\omega\!+\!\omega')
\label{eq:Sigma}
\end{equation}
where  $G(\vec k, \omega)$ is the large $N$ nematic mode propagator of Eq.~\eqref{eq:G}.  Notice that $\Sigma^{(i)}(\vec{p},\omega)$ have explicit non-trivial dependences  
on $\vec p$ and $\omega$. This momentum dependence makes it challenging to obtain the self-energy in closed form in general. Hence we will evaluate the self-energy and the resulting spectral function numerically. Nonetheless some valuable analytic understanding can be obtained in limiting cases. 
\begin{figure*}[hbt]
\parbox{.6\textwidth}{
\subfigure[$A(\vec{p},-9meV)$]{\includegraphics[width=.45\textwidth]{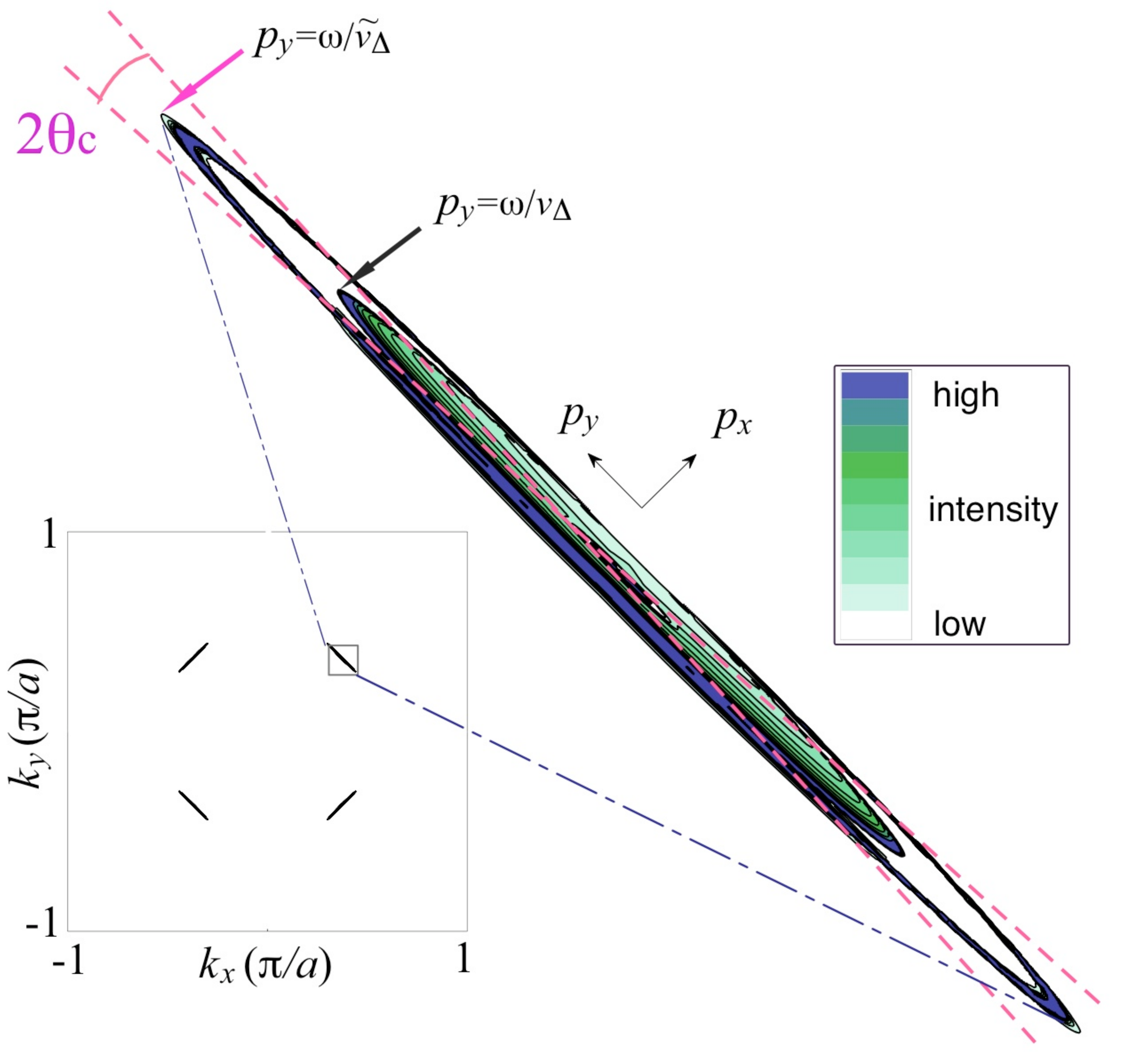} }
}
\parbox{.35\textwidth}{
\subfigure[]{\centering
\includegraphics[height=0.2\textheight]{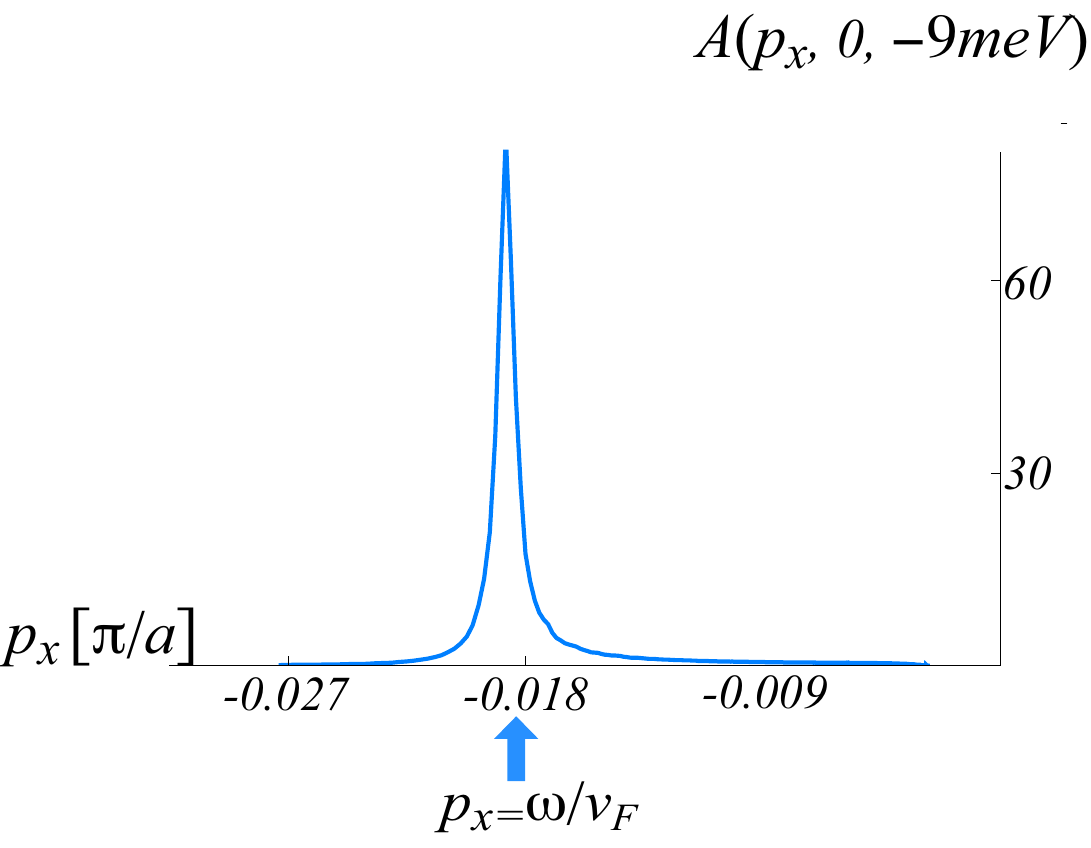}
}\\
\subfigure[]{\centering
\includegraphics[height=0.18\textheight]{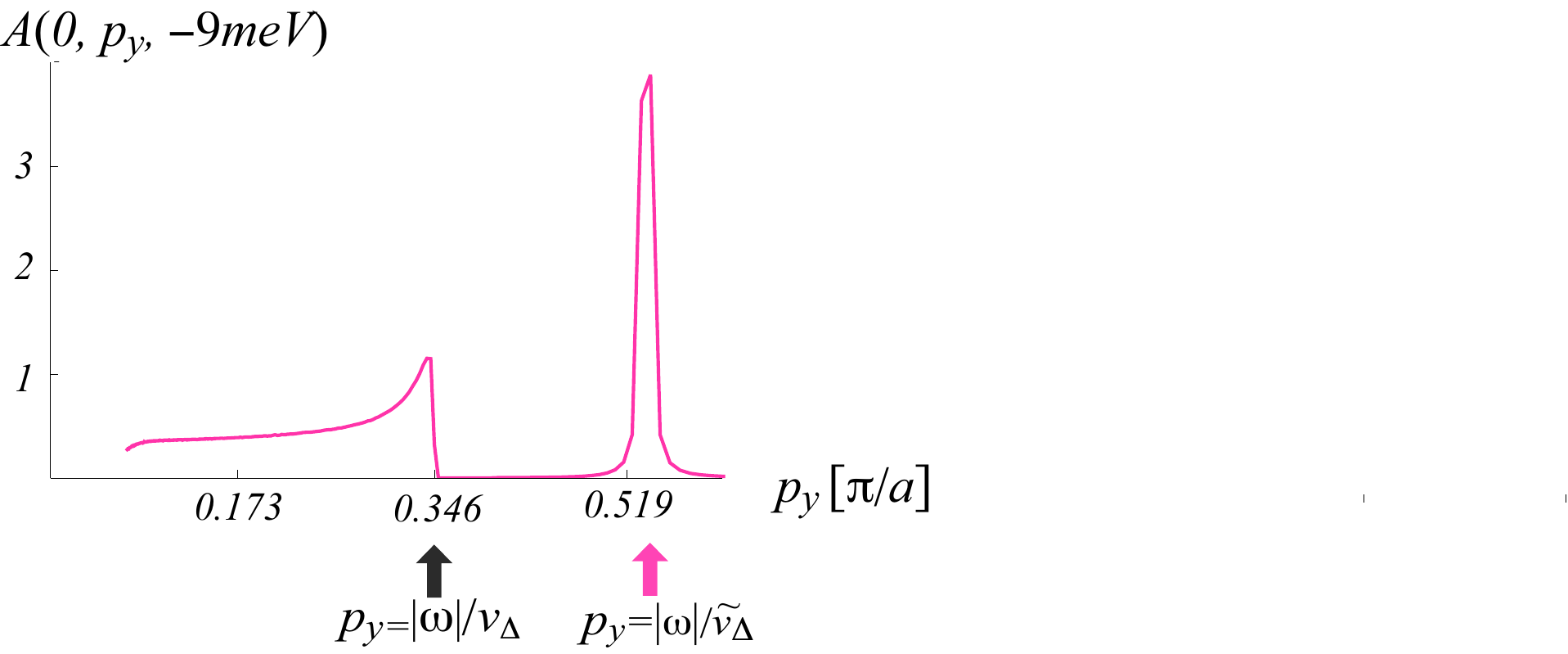}
}
}
\caption{Momentum distribution of the nodal nematic QCP
spectral function at a energy $\omega\!=\!-9meV$ for $v_F\!=\!0.508 eV(\pi/a)^{-1}$, $v_\Delta\!=\! 0.026eV(\pi/a)^{-1}$, $(v_F/v_\Delta\!=\!19.5)$.
Here, the momentum $\vec{p}$ is measured with respect to nodal point
$\vec{K}_1\!=\!(K,K)$ in a (rotated) local coordinate system 
so that $p_x$ and $p_y$ lie, respectively, along the the nodal and the tangential directions. 
(a) Contour plot (color online).    
$\theta_c$ is the critical angle at which a well defined qp peak 
departs from the incoherent continuum. For $\theta<\theta_c$, the qp is well defined. 
(b) and (c) are line cuts along the nodal direction and tangential direction respectively.
}\label{fig:MDC}
\end{figure*}
 First of all, in the limit of $\vec{p}=0$ (nodal point), we can show that the self-energy takes the form \cite{eakim-preprint}: 
\begin{equation}
\hat{\Sigma}(\vec{p}=0,\omega)\propto-\frac{\omega}{N}f(\alpha)\left[\ln\left(\!\frac{\omega^2}{\Lambda^2}\!\right)+i\pi\right]\mathbb{I}, 
\label{eq:Sigma-iso}
\end{equation}
where $f(\alpha)$ is a function of dispersion anisotropy ratio $\alpha$.  Thus the  imaginary part of self energy grows with energy and the width of the quasiparticle peak is never sharper than the energy.

Another important observation we can make from Eqs.(\ref{eq:Sigma} - \ref{eq:Sigma-iso}) is that the real part of self energy renormalizes  $v_\Delta$ downwards without affecting $v_F$. This is a combined result of two aspects of the theory. For one thing, each component $\Sigma^{(i)}(\vec{p},\omega)$ scales with $\omega, v_Fp_x,v_\Delta p_y$ respectively for $i=0,1,2$ as it can be easily checked from \eqref{eq:Sigma-iso}. Secondly, $\Sigma^{(2)}$ enters the spectral function with the opposite sign from $\Sigma^{(0)}$ and $\Sigma^{(1)}$ due to the structure of the interaction vertex, which reflects the fact that the coupling to the nematic mode breaks Lorentz invariance. 
 Since there is a logarithmic divergence in this renormalization of $v_\Delta$, the anisotropy ratio 
 $\alpha$  is relevant at $\alpha=0$. Hence it is clear that one should consider the case of a anisotropic dispersion $\alpha\neq0$ from the beginning. 

It is noteworthy that for non-zero $\vec{p}$,
any amount of anisotropy $\alpha\neq0$ qualitatively changes 
the analytic structure of the nodal Fermion self-energy $\hat{\Sigma}(\vec{p},\omega)$
from that at $\alpha=0$.
This is due to the existence of two energy scales $E_{1}({\vec{p}}) \neq E_{2}({\vec{p}})$ (see Eq.\eqref{eq:E1E2} for the definition of these energy scales)
entering the nematic mode propagator Eq.\eqref{eq:G} (see Fig.~\ref{fig:B-spectral}).  Such change in the analytic structure  is not perturbatively accessible in $\alpha$ from the $\alpha=0$ case. 
Hence we have computed $\hat{\mathbf\Sigma}$ numerically  from Eq.~\eqref{eq:Sigma} for an arbitrary nonzero value of  $\alpha$, by obtaining  $\textrm{Im} \hat{\mathbf \Sigma}$ via a Monte Carlo integration, and from this  obtained $\textrm{Re} \hat{\mathbf\Sigma}$ by Kramers-Kronig. We verified this method against the analytic expression Eq.\eqref{eq:Sigma-iso} in the isotropic limit.

\begin{figure}[h]
\subfigure[]{
\includegraphics[height=.2\textheight]{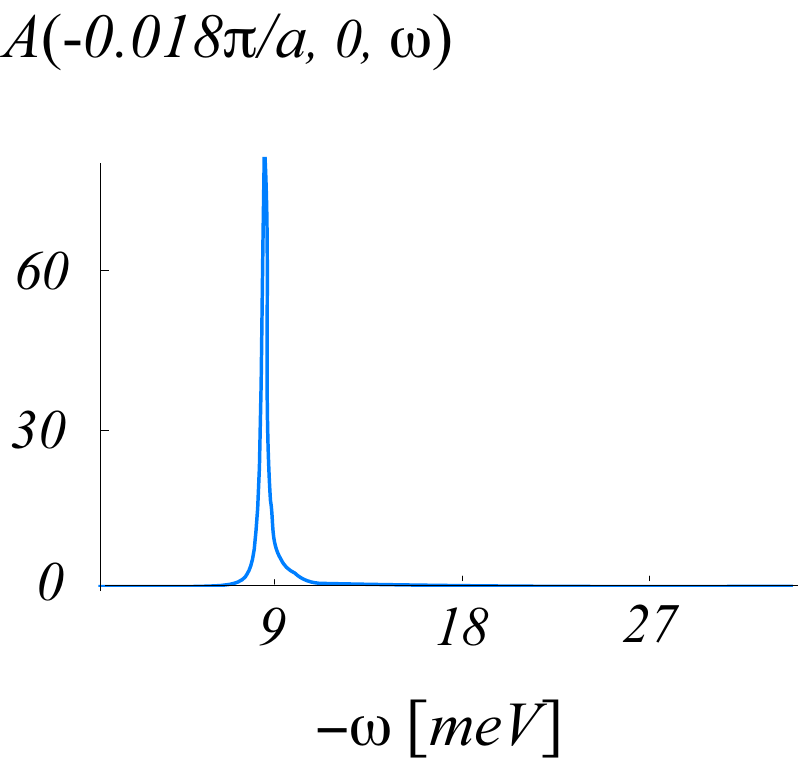}
}
\subfigure[]{
\includegraphics[height=.2\textheight]{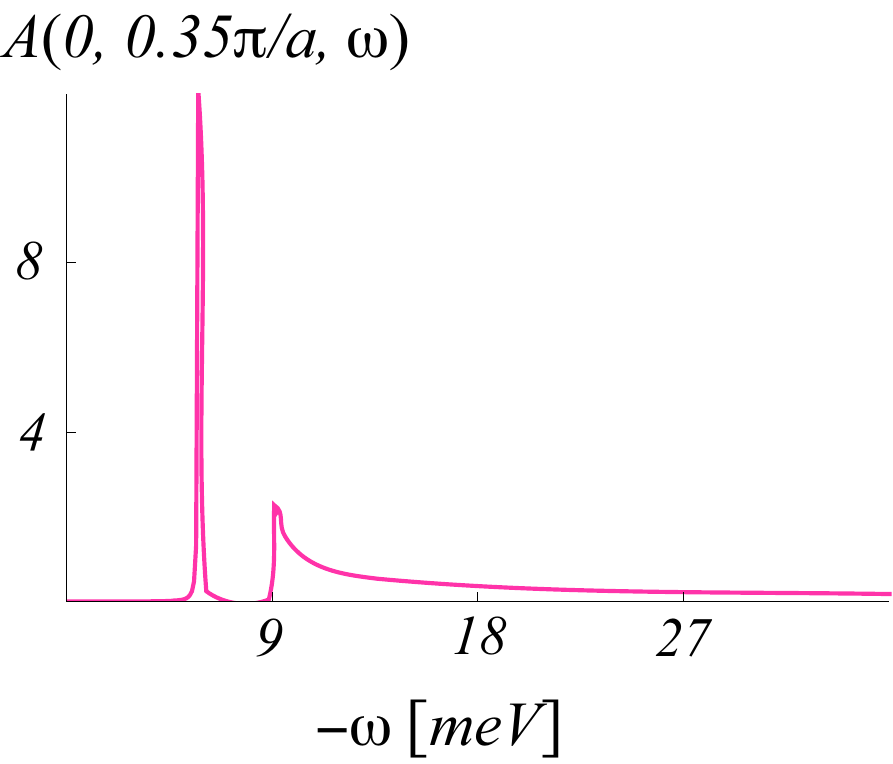}
}
\caption{EDC's taken at two points in momentum space. (a) At $\vec{p}=(-9meV/v_F,0)$. (b) At  
$\vec{p}=(0,9meV/v_\Delta)$. These are the momenta corresponding to the position of the peak in Fig.3(b) and to the threshold in Fig.3(c) respectively.}
\label{fig:EDC}
\end{figure}

We present the nodal nematic QCP fermion spectral function $A(\vec{p},\omega)$, obtained from the numerical calculation of $\hat{\bf\Sigma}(\vec{p},\omega)$, in Fig.\ref{fig:MDC} and Fig.\ref{fig:EDC}. 
Although we have carried out the calculation for arbitrary values of velocities, we used 
the dispersion $v_F= 0.508 eV(\pi/a)^{-1}$ and $v_\Delta=0.026 eV(\pi/a)^{-1}$ with the velocity ratio $v_F/v_\Delta=19.5$ in our plots in order the demonstrate the effect of a large bare velocity anisotropy. This dispersion was obtained by linearizing the phenomenological model Hamiltonian for BSCCO 
of M. Norman et al.\cite{norman}. For the value of the critical coupling, we used $\lambda_c^2/2N\gamma=0.3$. 
Fig.\ref{fig:MDC} shows 
the momentum distribution of the spectral function at a fixed energy $\omega=-9meV$.  The eccentricity of  the ellipses in the contour plot reflects the large bare velocity anisotropy. We also show two line cuts or momentum distribution curves (MDC's): one along the nodal direction and the other along the  tangential direction. In Fig.\ref{fig:EDC} we show representative energy distribution curves (EDC's) at two fixed points in momentum space (one point along the nodal direction, another point along the tangential direction).   

On the basis of this numerical calculation of $\hat{\Sigma}(\vec{p},\omega)$ and hence the spectral function $A(\vec{p},\omega)$, 
we extract two principal
qualitative effects of the nematic critical fluctuations on the nodal Fermion properties: (i) strongly momentum (angle) dependent scattering; (ii) an anisotropic renormalization of the already anisotropic bare velocities. 

The presence of strongly angle-dependent scattering is clearly shown in Fig.\ref{fig:MDC}(a).
The qp's are highly damped in the direction normal to the FS ($p_x$ direction) while they remain sharply defined in the vicinity of the FS ($p_y$ direction). More specifically, 
the momentum distribution of the spectral intensity 
exhibits distinct behaviors in different wedges of $p$-space,
which open about the FS at a critical angle, $\theta_c=
\tan^{-1}(v_\Delta/v_F)$.
For $\theta>\theta_c$, the peaks in the spectral function are 
 broad with  a width $\propto |\omega|$.
For $\theta<\theta_c$, the MDC has sharp peaks (the thin white lines) dispersing with a renormalized velocity $\tilde v_\Delta$,
enhancing the eccentricity of the constant energy ellipse. 
We have marked the boundary between different wedges with the dashed (pink) lines in Fig.\ref{fig:MDC}(a). Due to the extreme bare anisotropy, the angle $2\theta_c$ which defines the region in $\vec{p}$-space with well defined qp's is rather small. However, this narrow wedge can qualitatively affect the long time properties of the nodal fermions\cite{kim-lawler}.

The anisotropic renormalization of the dispersion  is particularly evident in the Fig.\ref{fig:MDC} (b) and (c).
The free nodal fermion theory would have placed sharply defined peaks at
$\vec p = (|w|/v_F,0)$ and 
$\vec p = (0,|w|/v_\Delta)$, in Fig.\ref{fig:MDC} (b) and (c) respectively.
Fig.\ref{fig:MDC}(b) indeed shows a peak (albeit broad) at the position expected from the bare value of the $v_F$ and hence $v_F$ is unrenormalized by $1/N$ fluctuations. However, the position of a sharp peak in Fig.\ref{fig:MDC}(c) has been shifted away from the bare position.
This shift is characterized by a renormalization of $v_\Delta \to \tilde{v}_\Delta$, where $\tilde{v}_\Delta$ is related to $v_\Delta$ by
\begin{equation}
\tilde{v}_\Delta=v_\Delta\left[1-\frac{1}{N}\Gamma\left(\frac{v_F}{v_\Delta}\right) -O\left(\frac{1}{N^2}\right)\right].
\label{eq:vD}
\end{equation}
Here $\Gamma(v_F/v_\Delta)$ is a positive function of $v_F/v_\Delta$. Hence
critical fluctuations effectively enhance the dispersion anisotropy. 
Such anisotropic renormalization of the dispersion is the result of the structure of the coupling between the nematic mode and the nodal fermions and it holds order by order in the $(1/N)$ expansion.
The consequences of a RG analysis of this decrease in $v_\Delta$ will be presented in a forthcoming paper\cite{huh}.

One can readily understand the kinematic origins of the sharply defined qp's near the FS inside the narrow wedge in Fig.~\ref{fig:MDC} (a). 
$E_1(\vec{p})$  defines the bare dispersion of a qp near the nodal point $\vec{K}_1=(K,K)$ while $E_2(\vec{p})$ defines the bare dispersion of a qp near the nodal point $\vec{K}_2=(K,-K)$.
Since the coupling to qp's near {\it both nodes} determines the dynamics of the nematic mode, it is the lesser of $E_1(\vec{p})$ and $ E_2(\vec{p}) $
that sets the threshold for decay.
For $\vec p = (p_x,0)=(-9meV/v_F,0)$ normal to the FS (Fig.~\ref{fig:EDC}(a)),
$E_2(\vec p)=v_\Delta |p_x| \ll E_1(\vec p) = v_F |p_x|$, so the qp is highly damped. Notice the asymmetric line shape which reflects  the $\vec{p}$-dependence of the self-energy.
For $\vec p=(0,p_y)=(0,9meV/v_\Delta)$ along the FS (Fig.~\ref{fig:EDC}(b)), the nematic fluctuations renormalize the qp velocity 
$v_\Delta \to \tilde v_\Delta$, and the qp energy, $\tilde E_1(\vec p)=\tilde v_\Delta |p_y| < E_1(\vec p)\ll E_2(\vec p)$, so there no damping.
It also produces an EDC with a `peak-dip-hump'. 
The existence of {\it entirely} undamped quasiparticles inside sharply defined $k$-space wedges is likely
an artifact of the first order corrections in the $1/N$ expansion, since higher orders terms are likely to introduce finite damping. It is therefore more plausible that the wedge delineates a crossover from a regime in which the qp's are highly damped, to a regime inside the wedge in which the qp peaks are relatively narrow with a width that vanishes as one approaches the tangential direction along the FS.

\section{The nematic glass} 
\label{sec:glass}

Consider now the nodal nematic phase away from criticality, $\lambda>\lambda_c$, where the nematic order parameter has a non-zero expectation value, as in Eq.\eqref{OP}.  Because of the Ising character of the ordered state, there is a gap in the nodal nematic fluctuation spectrum,  $\omega_{N}\sim T_N \sim (\lambda_c-\lambda)$.  At energies large compared to $\omega_N$, the spectral function is little different from its behavior at criticality.  However, at energies small compared to $\omega_N$, the nodal
qp's  are undamped, but the nodal positions are shifted from the symmetric points in $k$-space  by  an amount $\Delta p \propto \sqrt{\lambda-\lambda_c}$ as shown in 
Fig.~\ref{fig:PD}, and the velocity tangential to the Fermi surface is renormalized, as in Eq.\eqref{eq:vD}.  

Unfortunately, quenched  disorder (impurities)
has a devastating effect on nematic phases,  and indeed macroscopic manifestations of electron nematic order have only been seen in ultra-pure systems~\cite{Borzi2007,Cooper2002}. Specifically, the order parameter theory in the presence of impurities is equivalent to the random-field Ising model\cite{carlson,nattermann-97}, and hence nematic systems are more generally expected to exhibit glassy dynamics rather than a broken symmetry.\cite{dahmen-review}   Thus, more detailed spectroscopies and/or local measurements play an especially central role in experimental studies of nematic phases.   
Here we sketch the effects of disorder on the single-particle spectral function in the nematic phases. 
We consider two extremal cases:  
short range and mesoscopic range disorder.  

Short range disorder produces an imaginary part to the self-energy, which smears the MDC at $\omega=0$ in {\it all directions} about the nodal point. Due to the  intrinsic dispersion anisotropy, this broadened weight is severely elongated along the tangential direction so the resulting MDC has
 a shape that resembles a Fermi ``arc.'' However the width of the MDC in the normal direction  is still proportional to the arc length. 

In sufficiently clean systems, the effects of disorder are only significant on longer distances, producing a ``nematic glass''.
Here we focus on the  static effects leaving out (important) dynamical effects 
generally expected in glassy phases, such as hysteresis and noise.
If each domain is large enough to support a  defined local nematic order, the variance  $\sigma^2\equiv\overline{|\langle{\phi}\rangle|^2}$ among domains amounts to a distribution of nodal positions along the FS. In such a mesoscale glass phase, a local measurement (such as STM) will see sharply defined nodal quasiparticles at low enough energy, but a measurement, such as ARPES, which averages over a large area, will measure the domain averaged spectral function
\begin{equation}
\overline{A}(\vec p,\omega) = \int\!\! d\phi\; P(\phi)\; A_0(p_x,\; p_y\!+\!\phi/v_\Delta,\;\omega)
\label{eq:def-barA}
\end{equation}
where $P(\phi)$ 
is the probability density that $\phi$ which shifts the position of the node by $\Delta p_y=\phi/v_\Delta$, and $A_0(p_x,p_y,\omega)$ is the spectral function for free qp's
\begin{multline}
A_0(p_x,p_y,\omega)=\frac{\pi}{E_1(\vec{p})}\Big[\delta(\omega-E_1(\vec{p}))(E_1(\vec{p})+v_Fp_x)\\
+\delta(\omega+E_1(\vec{p}))(E_1(\vec{p})-v_Fp_x)
\Big]
\end{multline}
with $E_1(\vec{p})$ given in Eq.~\eqref{eq:E1E2}.
For a Gaussian distribution of $\phi$ with variance 
$\sigma$, $P(\phi)=e^{-\phi^2/2\sigma^2}/{\sqrt{2\pi}\sigma}$ and it is straight forward to show that the average spectral function in Eq.\eqref{eq:def-barA} becomes
\begin{multline}
\overline{A}(\vec p, \omega) 
=\sqrt{\frac{\pi}{2\sigma^2}}\,\Theta\left(\omega^2-(v_Fp_x)^2\right)\frac{|\omega|}{\sqrt{\omega^2-(v_Fp_x)^2}}\\
\ \times \left[e^{-\frac{\left(v_\Delta p_y-\sqrt{\omega^2-(v_Fp_x)^2}\right)^2}{2 \sigma^2}}+e^{-\frac{\left(v_\Delta p_y+\sqrt{\omega^2-(v_Fp_x)^2}\right)^2}{2 \sigma^2}}
\right]
\label{eq:glassA}
\end{multline}
Note that this defines a Fermi arc of length $\sigma/v_\Delta$ with a vanishing perpendicular width,
along the FS ($p_x=0$) at zero energy $\omega=0$. 

\section{Discussion}
\label{sec:discussion}

In this paper we have presented a phenomenological theory of 
the nodal nematic QPT. 
Formally similar problems arise in studies of the nodal quasiparticles near other QPTs.\cite{Vojta2000prl,Vafek2002,hermele2005}
In all these cases, the ordered phase preserves the $C_{4v}$ symmetry, and 
the velocity anisotropy is irrelevant so the fixed point is isotropic with emergent Lorenz invariance.   In the case of a nodal nematic critical point, the dispersion anisotropy is enhanced at criticality. This has unforeseen effects on the single-particle spectral function.

We conclude with a few observations  
on the possible 
relevance of a nodal nematic quantum criticality to the cuprate physics, recognizing the danger of extrapolating large $N$ results down to $N=2$\cite{huh}.
The previously cited evidence of a nematic phase YBa$_2$Cu$_3$O$_ {y}$ near $y\sim6.5$
from\cite{Ando2002} transport anisotropy measurements
and from inelastic neutron scattering in untwined
samples\cite{hinkov06}, involves studies of materials with unusual purity and crystalline perfection.
Nematic order couples linearly  to spatial inhomogeneity and disorder, which thus  generically causes the ordered phase to be replaced  by a glassy (domain) phase.
A glassy phase with anisotropic domains has been seen in 
STM studies of underdoped {Bi$_2$Sr$_2$CaCu$_2$O$_{8+\delta}$}\cite{Kohsaka2007}.
It is tempting to identify these structures, and the ``Fermi arcs'' seen in ARPES experiments on the same material\cite{Kanigel2006}, with the nematic glass.  
However, this interpretation requires an extrapolation from the phase with strong  SC order
where the theory is valid to the non-superconducting pseudogap regime where the experiment is carried out. Moreover, this interpretation is far from unique.\cite{varma,paramekanti,berg}
A critical test of this interpretation is that as the disorder is made increasingly weak, there will be a crossover from a short range disorder behavior with a nodal point that is broadened in all direction,  to a nematic glass regime 
with a longer $T=0$ arc length but a narrower width 
 (with Eq.\eqref{eq:glassA} as the limiting case).
 The existence of a nodal nematic phase\cite{valla2006}, related to ordered or fluctuating stripes, is moderately clear\cite{kivelson2003} in La$_{2-x}$Sr$_x$CuO$_4$ and related materials near $x=1/8$, but it seems likely that the critical phenomena in this system will be considerably affected by quenched disorder. 

The structure of the single particle spectral function at the nodal nematic QCP shown in Fig.\ref{fig:MDC} 
with sharp qp's within a narrow wedge around the FS
but otherwise broad, may provide  a consistent explanation for both the observation of broad qp's in ARPES\cite{zxrmp03} and of sharp qp interference peaks in STM\cite{mcelroy03}\cite{dhlee} on \BSCCO.
Since the wedge is so narrow due to the extreme dispersion anisotropy, it is unlikely to be observed in a direct momentum space probe.  Hence in ARPES, one is likely to only observe  broad qp's. 
However the interpretation of peaks in the Fourier transform of STM on \BSCCO\ in terms of 
qp interference invoke the notion of coherent qp's at the tips of the equal energy ellipses (or ``banana''s as it is frequently referred to) along the FS\cite{mcelroy03, dhlee}. 
The narrow wedge provides a natural and possibly unique mechanism for this interpretation.
This aspect will be further analyzed in a future publication\cite{kim-lawler}.

\noindent{\bf Acknowledgments}
We thank E.\ Berg,  A.\ Chubukov, H.-Y.\ Kee, M.\ Norman, A.\ Paramekanti, D.J.\ Scalapino, J.\ Tranquada, O.\ Vafek, M.\ Vojta, E.\ Zhao for discussions.
 This work was supported in part by the NSF, under grants DMR 0442537(EF) and DMR 0531196(SAK), by the DOE under contracts DE-FG02-91ER45439(EF) and 
DE-FG02-06ER46287(SAK), by the Stanford Institute for Theoretical Physics(EAK), by the CRC(MJL), and by the Urbanek Family Fellowship(PO).


\begin{thebibliography}{99}
\bibitem{huh}
In an upcoming paper of Y.\ Huh and S.\ Sachdev, the nature of the fixed point to order $1/N$ is studied, and it is found that there are logarithmically slow flows to infinite anisotropy. Moreover, it is found that large anisotropy serves the same role in the theory as large $N$, thus giving corroborating evidence that the large $N$ results may be applied to the physical case of $N=2$.
\bibitem{order-of-limits}
  These two results are not necessarily in contradiction with each other. What they mean is that the quantum phase transition is first order for
  small $\epsilon$  and finite $N$, while it is continuous at large $N$ and fixed dimension. In the $\epsilon-N$ plane the first order and continuous phase transition regions should then coalesce as $\epsilon \to 0$ and $N \to \infty$, and the order of limits does not commute.
\end{thebibliography}
 \end{document}